\documentstyle[titlepage,12pt]{article}
\begin{document}
\parindent 2em
\baselineskip 4.5ex

\begin{titlepage}
\begin{center}
\vspace{12mm}
{\LARGE Critical Fluctuations in Superconductors 
and the Magnetic Field Penetration Depth}
\vspace{15mm}

Igor F. Herbut$^{a}$ and Zlatko Te\v sanovi\' c$^{b}$  \\

$^{a}$Department of Physics and Astronomy, University of British Columbia, 
6224 Agricultural Road,\\
Vancouver B. C., Canada V6T 1Z1\\
$^{b}$Department of Physics and Astronomy, Johns Hopkins University, 
Baltimore, 
MD 21218, USA

\end{center}
\vspace{10mm}
{\bf Abstract:}
The superconducting transition is studied within the one-loop RG
in fixed dimension $D=3$ and at the critical point. 
A tricritical behavior is found and, for $\kappa > \kappa_c$,
an attractive charged fixed point, distinct from that of a
neutral superfluid.  The critical exponents of the continuous
transition are evaluated and
it is shown that the anomalous dimension of 
the gauge-field equals unity. This implies the
proportionality of the magnetic field 
penetration depth and the superconducting
correlation length below the transition. 
The penetration depth exponent
is non-classical. We argue that it cannot be extracted from
the dual theory in a straigthforward manner since it is not
renormalized by fluctuations of the dual field.


\end{titlepage}


The problem of a charged scalar field coupled to a gauge vector potential 
arises frequently in theoretical physics. 
In its original version, it describes
formation of Meissner state in 
superconductors \cite{halperin} and Higgs mechanism in particle
physics \cite{coleman}. Furthermore, the nematic-smectic-A 
transition in liquid crystals
\cite{degennes} and, more recently, the 
transitions between plateaus in the quantum Hall effect
\cite{wen,pryadko}
and the finite-magnetic-field critical behavior in
extreme type-II superconductors \cite{tesa} have also
been related to this problem. 
In a superconductor, the scalar field 
represents the fluctuating superconducting order
parameter $\Psi$ which, BCS pairs being charged, is coupled to fluctuations
in the electromagnetic potential $\vec{A}$. At the mean-field level the
transition is discontinuous and remains so when fluctuations in $\Psi$
are included via the $\epsilon (\equiv 4-D)$-expansion \cite{halperin}. 
Numerical simulations 
\cite{das,bartho} of related lattice models support this scenario 
for smaller values
of the Ginzburg parameter $\kappa$.
However, for large $\kappa$, the results are consistent with a
continuous, second-order phase transition \cite{das,bartho}. 
The picture obtained in
numerical work is in accordance with the $1/n$-expansion \cite{halperin,leo} 
and with duality arguments which
connect the lattice version of the theory to a dual gas of interacting
vortex loops and the ``inverted'' 3D XY model \cite{das,kleinert}.
 
In this Letter we study this superconducting-Higgs electrodynamics
(SHE) directly in $D=3$ within perturbation theory
at the critical point corresponding to the charged
superfluid. Our results are as follows:
We first show that the anomalous dimension
of the gauge field, $\eta_A$, equals unity to all orders in perturbative
expansion. By combining this result with the Josephson relation,
we argue that the magnetic field penetration depth, $\lambda$,
and the superconducting correlation length, $\xi$,
diverge with the same exponent ($\nu$) as the transition is approached
from below. These results should be contrasted
with $\eta_A = 0$ and $\lambda\propto \sqrt{\xi}$ at the {\em unstable}
critical point for neutral superfluid. We then demonstrate that
our one-loop results imply the presence of the tricritical point
in SHE. For bare $\kappa < \kappa_c$ the renormalization group (RG) 
flows are unstable and
the transition is likely first order while,
for $\kappa > \kappa_c$, we find a stable fixed point 
indicating a continuous
transition. By selecting the regularization which reproduces
the established numerical value of $\kappa_c$, we evaluate $\nu$ and 
the anomalous dimension of $\Psi$, $\eta_{\Psi}$.  
Our results strongly suggest that
the exponent for $\lambda$ is non-trivial and therefore different from
the mean-field value suggested by Kleinert et al. \cite{kleinert} 
We propose that this difference has a physical 
origin and that this form of 
dual theory does not offer any simple way of determining
the exponent of $\lambda$. Since the 
penetration depth is of direct experimental interest
\cite{hardy,goldman}, the description
of the critical behavior within the {\em original} SHE theory 
retains its physical significance.

We are interested in the infrared behavior of the SHE:
\begin{equation}
H=\int d^{D}\vec{r} [|(\nabla - i e \vec{A})\Psi(\vec{r})|^{2} + 
\mu^{2} |\Psi(\vec{r})|^{2} + \frac{b}{2}|\Psi(\vec{r})|^{4}+ 
\frac{1}{2}(\nabla \times\vec{A})^{2}]~~~,    
\end{equation}
where $\mu^{2}\propto (T-T_{c0})$, $T_{c0}$ is the mean-field transition 
temperature, $b$ is a temperature independent 
constant, and $e$ is the charge of a BCS pair. 
For generality, we assume that the order parameter 
$\Psi$ has $n$ complex components and that the system is 
$D$-dimensional, $n=1$ and $D=3$ being eventually the case 
of physical interest. 
We chose to work in the gauge where the vector potential 
is purely transverse, i. e. where the bare gauge-field propagator is:
$D_{ij}(\vec{q})=(\delta_{ij}-\hat{q}_{i}\hat{q}_{j})/q^{2}$. 
First, let us discuss the non-perturbative results concerning 
the anomalous dimension of the gauge-field and the divergence of the 
penetration depth.
The anomalous dimension of the gauge-field propagator is defined as:
\begin{equation} 
\eta_{A}=-\lim_{p\rightarrow 0}\frac{d\log Z_{A}}{d\log(p)}
\end{equation}
where $Z_{A}$ is the gauge-field renormalization factor and $p$ is 
the momentum of the gauge-field propagator. The $\beta$-function for the 
charge is 
\begin{equation}
\beta_{e}=\frac{d\hat{e}^{2}_{r}}{d\log(p)}=\hat{e}^{2}_{r}
(D-4-\frac{d\log Z_{A}}
{d\log(p)})~~~, 
\end{equation}
where $\hat{e}^{2}_{r}=e^{2}_{r}/p$ is 
the dimensionless renormalized charge. 
On approaching the critical point, $\hat{e}_{r}\rightarrow\hat{e}_{0}$ and 
$\beta_{e}\rightarrow 0$, so taking the limit $p\rightarrow 0$ in Eq. (3)
we obtain:
\begin{equation}
\hat{e}_{0}^{2}(D-4+\eta_{A})=0~~~. 
\end{equation}
Assuming a stable charged fixed point in the theory $\hat{e}_{0}\neq 0$ 
one obtains 
\begin{equation}
\eta_{A}=4-D~~~. 
\end{equation}
This exact result has an important physical 
consequence, since it modifies the relation between the correlation 
length and the penetration depth as the critical point is approached from 
below. In general, the penetration depth scales with the superfluid 
density below the transition as:
\begin{equation}
\lambda^{\eta_{A}-2}\propto\rho_{s}~~~, 
\end{equation}
where $\rho_{s}\propto\xi^{2-D}$ 
is the Josephson relation \cite{joseph}. 
Close to the transition controlled 
by the attractive charged fixed point it follows that
\begin{equation}
\lambda\propto\xi
\end{equation}
for all $D$. In contrast to the scaling governed by the XY fixed point,
where $\eta_{A}=0$ and $\lambda\propto\xi^{(D-2)/2}$, 
the ratio between the two lengths close to the charged critical point  
approaches a finite constant. The divergences of both 
lengths are determined by the same exponent $\nu$.  

To obtain the flow-diagram for the coupling constants 
and the values of critical exponents one must 
rely on some approximation for the $\beta$-functions. 
Here we perform the 
perturbative calculation of $\beta$-functions in fixed dimension and 
at the critical point. 
To the lowest order in perturbation theory, 
the contributions to the self-energy, 
polarization and the quartic vertex are given by the 
diagrams in Fig. 1. 
Due to the choice of gauge, 
the remaining one-loop diagrams for the quartic vertex all vanish when 
the external momenta go to zero \cite{coleman}. Note that this
procedure explicitly preserves Ward identities associated with gauge
invariance. Since we wish to work directly in 
$D=3$, we are forced to define the renormalized value of $b$ at a 
finite momenta of external legs to avoid the infrared 
divergence in the last diagram on Fig. 1. 
This divergence is a consequence of 
gauge invariance which requires massless gauge-field.
The renormalized 
coupling constant $b_{r}$ is defined at the usual symmetric point: 
\begin{equation}
\vec{k}_{i}\cdot\vec{k}_{j}=\frac{4\delta_{ij}-1}{4}p^{2},~~~ i,j=1,2,3
\end{equation}
and at the critical point where the renormalized mass of $\Psi$ vanishes.
There are two  relevant coupling constants in the problem: the quartic 
term coupling and the charge. The 
standard procedure \cite{zinn} gives the renormalized coupling 
constants to lowest order:
\begin{equation}
e^{2}_{r}=e^{2} + \frac{2n\Gamma(1-D/2)\Gamma^{2}(D/2)}{ (4\pi)^{D/2}
\Gamma(D)} e^{4} q^{D-4}~~~, 
\end{equation}
\begin{eqnarray*}
b_{r}=b-( \frac{n+3}{2^{2-D/2}}+1 )\frac{\Gamma(2-D/2)
\Gamma^{2}(D/2-1)}{(4\pi)^{D/2} \Gamma(D-2)}b^{2} p^{D-4} +\\
\frac{4(D-1) \Gamma(2-D/2)\Gamma(D/2-1)\Gamma(D/2)}{(4\pi)^{D/2} 
\Gamma(D-1)} b e^{2} p^{D-4} - \frac{2(D-1)\Gamma(2-D/2) 
\Gamma^{2}(D/2-1)}{(4\pi)^{D/2} \Gamma(D-2) 2^{2-D/2}}e^{4}p^{D-4},  
\end{eqnarray*}
where $\Gamma(x)$ is the factorial function, the momentum scale $p$ 
is defined above and $q$ is the momentum of the gauge-field propagator. 
Hereafter, we set $D=3$ and  $n=1$ and define dimensionless couplings with 
respect to the external momentum $p$: $\hat{b}=b/p$, $\hat{e}^{2}=e^{2}/p$.
If we choose the reference 
momentum of the gauge-field propagator as $q=p/c$, where c 
is a constant, the $\beta$-functions are: 
\begin{equation}
\beta_{e}(\hat{b}, \hat{e})\equiv\frac{d\hat{e}^2}{d\log(p)}
=-\hat{e}^{2}+\frac{c}{16}
\hat{e}^{4}~~~, 
\end{equation}
\begin{equation}
\beta_{b}(\hat{b},\hat{e})\equiv\frac{d\hat{b}}{d\log(p)}
=-\hat{b} + \frac{1}{8}(2 \sqrt{2}+
1)\hat{b}^{2}-\frac{1}{2}
\hat{b} \hat{e}^{2} + \frac{1}{2 \sqrt{2}}\hat{e}^{4}~~~,  
\end{equation}
and we dropped the subscript ``$r$'' for renormalized coupling constants 
in the last two equations. 

The above $\beta$-functions explicitly depend on the 
introduced ratio of momenta, $c$. This reflects the known property of 
the RG in fixed dimension that, unlike in the $\epsilon$-expansion scheme, 
the coefficients in the perturbative series for
 $\beta$-functions are procedure 
dependent \cite{zinn}. Consequently, an additional 
information is needed to fix the value of $c$ in the 
one-loop calculation. We now note an important property of Eqs. (10,11):
If $c$ is treated as a free parameter, the RG flow diagram has
a structure pictured in Fig. 2 (for $c> 5.16$).
 Besides the standard Gaussian 
and the neutral superfluid fixed points, there are two charged fixed points 
of the above flow equations:
$(\hat{b}_{-},\hat{e}^{2}_{0})$
 and $(\hat{b}_{+},\hat{e}^{2}_{0})$, where $\hat{e}^{2}_{0}=16/c$ 
and $\hat{b}_{+}$ and $\hat{b}_{-}$ are the real roots 
of the equation $\beta_{b}(\hat{b},\hat{e}_{0})=0$. 
Stability analysis shows that the fixed point 
with the larger value of $\hat{b}=\hat{b}_{+}$ 
is attractive, while the one with $\hat{b}=\hat{b}_{-}$
is unstable in the direction of quartic term coupling. 
Now we look for the 
straight RG trajectories, $\hat{b}=2\kappa^{2}\hat{e}^{2}$, 
by demanding the invariance of the constant $\kappa$ under RG 
transformation:
\begin{equation}
\frac{d(\hat{b}/\hat{e}^{2})}{d\log(p)}=0~~~. 
\end{equation}
In our one-loop analysis this leads to the equation:
\begin{equation}
\beta_{b}(2\kappa^{2}\hat{e}^{2}_{0},\hat{e}^{2}_{0})=0~~~,
\end{equation}
resulting in two straight-line RG trajectories for which
\begin{equation}
\kappa^{2}_{+,-}=\frac{\hat{b}_{+,-}}{2\hat{e}^{2}_{0}}~~~. 
\end{equation}
Together with the result for the 
stability of the $(\hat{b}_{-},
\hat{e}^{2}_{0})$ fixed point, this implies that there is a tricritical 
line in the theory given by a GL parameter:
\begin{equation}
\kappa^{2}_{c}=\frac{ \hat{b}_{-} }{ 2\hat{e}^{2}_{0}}=
\frac{c+8-\sqrt{c^{2}+16 c-32(2+\sqrt{2})}}{8(2\sqrt{2}+1)}~~~.  
\end{equation}
This solution appears physically plausible and 
is in accord with Ref. \cite{kleinert1} 
in that the tricritical point is determined by a 
particular value of the Ginzburg parameter, as one would expect from a 
mean-field argument. Consequently, we 
fix the value of $c$ by demanding
that $\kappa_{c}$ in Eq. (15) matches the value 
obtained via the duality transformation of 
the lattice SHE \cite{kleinert1} 
(see also numerical Monte Carlo results 
of Ref. \cite{bartho}): $\kappa_{c}\cong 0.8/\sqrt{2}$. 
This condition results in $c=5.7$ ($> 5.16$). 
The other straight trajectory given 
by the $\kappa_{+}$ solution connects 
the Gaussian and the superconducting fixed point.

Having fixed the value of the parameter $c$ 
we thus obtain the flow diagram 
of Fig. 2 \cite{lawrie}, \cite{kovner}. 
The exponents at the attractive fixed point are 
\begin{equation}
\eta_{\Psi}=\frac{-\hat{e}_{0}^{2}}{4}=-0.70~~~,
\end{equation}
\begin{equation}
\nu=\frac{1}{2}( 1+\frac{\hat{b}_{+}}{8}-\frac{\hat{e}^{2}_{0}}{8} )
=0.53~~~, 
\end{equation}
with the numerical values calculated 
for $c=5.7$, ($\kappa_c = 0.8/\sqrt{2}$).  
It is worth mentioning that our procedure 
gives a respectable value for the 
correlation length exponent at the neutral superfluid fixed point: 
$\nu_{XY}=0.63$. 
The value of the anomalous dimension is rather large and negative 
but it does satisfy $\eta_{\Psi} >-1$ in $D=3$. 
Note that small reductions in the assumed value for $\kappa_c$
rapidly make $\eta_{\Psi}$ less negative (as do the next
order perturbative terms) while the value for $\nu$ is more robust.
For example, $\kappa_c=0.45/\sqrt{2}$ \cite{bartho} 
gives $\eta_{\Psi}=-0.20$ and 
$\nu=0.62$. The other exponents follow from standard scaling relations. 
It is conceivable, however, that the 
hyperscaling relation does not hold, due to the presence
of long-range gauge forces in (1). In that case 
we could define a characteristic dimension, $d_c$, from
$2-\alpha = d_c\nu$. By combining our results with the
prediction of the dual theory $\alpha =\alpha_{XY}=
-0.013$ one obtains $d_c \cong 3.8$. This is 
close to $d_c = 4$ which would arise from a naive scaling
of current-current interactions. 

To the lowest order in $\epsilon\equiv 4-D$ the 
$\beta$-functions derived from 
Eqs. (9) completely reproduce the results of the RG defined at zero 
external momenta and finite mass \cite{halperin,lawrie}. 
This is to be expected since the dimensional regularization and 
the minimal subtraction scheme 
lead to unique values of the coefficients in the 
$\beta$-functions \cite{zinn}. The parameter $c$ then does not 
appear at all, the attractive fixed point exists only 
for $n>182.95$ and the exponents agree with ours in this limit.
\cite{footnoteO}

The exponent for the magnetic field penetration 
depth is $\nu = 0.53 (0.62)$.
While this appears close to the mean field value of $1/2$ 
suggested by Kleinert et al. \cite{kleinert} on the basis
of the dual theory, it is clear from our procedure
that this exponent is non-trivial and determined by the structure
of the charged fixed point. This is an important aspect of the physics
of this problem: In the dual approach
the partition function of the original problem is related to the 
one for the interacting vortex loops \cite{das}. 
The dual description 
is a theory of a scalar complex field coupled to a {\em massive}
vector potential \cite{kleinert} and is in the universality class
of the 3D XY model \cite{kleinert}.  This description could be useful in 
providing the information 
on thermodynamic quantities. For example,
the specific heat exponent $\alpha$ {\em can} be calculated from the
hyperscaling relation, which holds in the dual theory, and
has the value for the 3D XY model. The dual approach, however,
offers little help in calculating exponents that 
characterize correlation functions of the original superconducting 
order parameter. In particular, 
the magnetic field penetration depth appears in the dual
description via the mass of the ``dual'' vector field.
If one assumes that this mass vanishes fast enough at the transition
at the {\em bare} level, then, at the critical point of the dual theory, 
this mass does not renormalize, 
i.e. it remains equal to its bare value.\cite{footnote} 
Since the mass of the dual vector field
is what determines the inverse of the penetration depth, the
implication is that the divergence 
of the penetration depth at the superconducting 
transition {\em cannot} be extracted from the RG analysis
of the dual model, even in principle. 
To calculate the penetration depth exponent it is necessary to
apply the RG analysis to the {\em original} SHE theory.
This fact supplied the primary motivation for the present work. 

Our value for the penetration depth exponent, $\nu =0.53$, 
is not seen in the experimental results of Ref. \cite{hardy},
which are consistent with the neutral 3D XY behavior,
but is in excellent agreement with those of Ref. \cite{goldman}.
While the critical region in which charge of BCS pairs is
relevant is rather narrow (even for HTS) 
due to the small value of the fine
structure constant, the fact that our exponent is quite close
to the mean-field value may effectively ``enlarge'' this region.
In general, as we approach the transition from below,
we expect that the mean-field behavior of the 
penetration depth crosses over to the intermediate regime
of neutral superfluid followed by the ultimate charged 
superfluid critical behavior. The exponent in this intermediate
3D XY regime is $\sim 1/3$, considerably different from
both the mean-field value and the ultimate value of $\sim 0.53$.
Consequently, on purely empirical grounds, 
the region over which the
penetration depth can be described by the crossover behavior
should be quite narrow. Finally, we should mention that 
experiments on other systems \cite{wen,pryadko,tesa} might
be even more promising in studying the SHE critical behavior
since there the effective ``fine structure constant'' can
be of order unity.

In summary, we studied the superconducting transition
by calculating the one-loop $\beta$-functions 
in fixed dimension $D=3$. The gauge invariance 
leads to infrared divergences which are handled by defining the 
renormalized coupling constants at finite values of the external momenta 
and right at the critical point. We eliminate the remaining freedom in the 
definition of renormalized couplings by requiring that 
our analysis yields to the
previously established \cite{kleinert1,bartho} 
numerical value of the Ginzburg
parameter which characterizes the
tricritical point. 
We then evaluate critical exponents at the attractive charged fixed
point of our theory.  It is shown that the 
penetration depth and the correlation 
length diverge in the same way close to the charged 
fixed point, to all orders in perturbation theory.\cite{germans} 

The authors acknowledge useful conversations with 
Professors I. Affleck and O. T. Valls and hospitality of the
Aspen Center for Physics where part of the work was
performed. This work has been supported in part by the NSF Grant No. 
DMR-9415549. IFH has also been supported by NSERC of Canada.

\pagebreak
 
Figure Captions:

Figure 1. Lowest order contributions to the self-energy (a), polarization 
 (b) and the quartic vertex (c). The full and dashed lines are the order 
parameter and the gauge-field propagators, respectively. 

Figure 2. The schematic flow-diagram for the dimensionless charge 
$\hat{e}$ and the dimensionless quartic term coupling $\hat{b}$ 
(for $c>5.16$).
Note that, for $c=5.7$, $\hat{e}_0^2 = 2.81$, $\hat{b}_+ = 3.22$,
$\hat{b}_- = 1.80$, $\kappa_+ = 1.07/\sqrt{2}$, $\kappa_- = 0.80/\sqrt{2}$.

\pagebreak


\begin{thebibliography}{99}
\bibitem{footnoteO}In this context we should note that within
our theory the numerical value for $c$ $(= 5.7)$ is sufficiently large 
to allow for charged critical points even in 3D (i.e., $\epsilon =1$) 
and for $n=1$.
\bibitem{footnote}Under the assumption that the penetration depth $\lambda$
goes as $(T_c - T)^{-\mu}$, our RG analysis of the dual theory shows that
$\lambda$ is not renormalized by the fluctuations 
of the {\em dual} field if $\mu > 1/3$. The mean-field exponent for
$\lambda$ suggested in Ref. \cite{kleinert} 
satisfies this condition (as does the value 0.53 found in
this work) but it is clear
from our analysis that this exponent cannot be determined solely from
the dual theory. The exponent for $\lambda$ {\em is} renormalized 
by the critical fluctuations of the {\em superconducting} order parameter
and takes on a non-trivial value when such fluctuations are accounted for
within the original SHE theory.
\end{thebibliography}
\end{document}